\documentclass[prl,groupedaddress,showpacs,twocolumn]{revtex4}
\usepackage[dvips]{graphicx}
\newcommand{\ped}[1]{\ensuremath{_{\rm #1}}}

\begin{document}

\title[Direct evidence for two-band superconductivity in MgB$\ped{2}$]
{Direct evidence for two-band superconductivity in MgB$_2$ single
crystals from directional point-contact spectroscopy in magnetic
fields}

\author{R.S. Gonnelli \email{E-mail:gonnelli@polito.it},
D. Daghero, G.A. Ummarino}

\affiliation{INFM - Dipartimento di Fisica, Politecnico di Torino,
Corso Duca degli Abruzzi 24, 10129 Torino, Italy}

\author{V.A. Stepanov}
\affiliation{P.N. Lebedev Physical Institute, Russian Academy of
Sciences, Leninski Pr. 53, 119991 Moscow, Russia}

\author{J. Jun, S.M. Kazakov and J. Karpinski}
\affiliation{Solid State Physics Laboratory, ETH, CH-8093
Z\"{u}rich, Switzerland}

\pacs{74.50.+r, 74.80.Fp, 74.70.Ad}

\begin{abstract}
We present the results of the first directional point-contact
spectroscopy experiments in high-quality MgB$\ped{2}$ single
crystals. Due to the directionality of the current injection into
the samples, the application of a magnetic field allowed us to
separate the contributions of the $\sigma$ and $\pi$ bands to the
total conductance of our point contacts. By using this technique,
we were able to obtain the temperature dependency of each gap
independent of the other. The consequent, strong reduction of the
error on the value of the gap amplitude as function of temperature
allows a stricter test of the predictions of the two-band model
for MgB$\ped{2}$.
\end{abstract}
\maketitle

During the last year, the consensus has been growing within the
scientific community on the fact that most of the features of
MgB$\ped{2}$ discovered so far can be properly explained by
admitting that two band systems are present in this new
superconductor: quasi-2D $\sigma$ bands arising from hybrid $sp^2$
orbitals in the boron planes, and 3D $\pi$ bands that stem from
the out-of-plane $p\ped{z}$ orbitals \cite{twobands,Choi}. The
unusual consequence of this band structure is that two different
energy gaps can be observed in clean limit
\cite{Choi,Liu,Brinkman}: $\Delta\ped{\sigma}$ (the larger) and
$\Delta\ped{\pi}$ (the smaller). Both gaps are expected to close
at the same temperature $T\ped{c}$ because of an inter-band
pair-scattering mechanism \cite{Suhl} but, while
$\Delta\ped{\sigma}(T)$ should approximately follow a BCS-like
curve, a marked reduction of $\Delta\ped{\pi}(T)$ with respect to
a BCS-like behavior is expected at $T\gtrsim$20~K
\cite{Liu,Brinkman}.

So far, one of the most convincing experimental supports of this
model has been the observation of two gaps by tunneling
\cite{tunnel} and point-contact spectroscopy \cite{Szabo} in
polycrystal samples and films. However, a direct and accurate test
of the predictions of the \makebox{two-band} model has been so far
impossible due to the lack of high-quality single crystals large
enough to be used for direction-controlled point-contact and
tunnel spectroscopy.

In this Letter, we present the results of the first directional
point-contact measurements in large single crystals of
MgB$\ped{2}$. We injected current along the $ab$ plane or along
the $c$ axis, and applied a magnetic field either parallel or
perpendicular to the $ab$ planes. This allowed us to separate the
partial contributions of the $\sigma$ and $\pi$ bands to the total
conductance, and to fit them obtaining the temperature dependency
of each gap with great accuracy. We will show that all the results
of this technique confirm very well the predictions of the
two-band model.

The high-quality MgB$_2$ single crystals used for our
point-contact experiments were produced at ETH
(Z$\mathrm{\ddot{u}}$rich) by starting from a mixture of Mg and B.
This mixture was put into a BN container and the crystals were
grown at a pressure of 30-35 kbar in a cubic anvil device. The
thermal process includes a one-hour heating up to
1700-1800$^{\circ}$C, a plateau of 1-3 hours, and a final cooling
lasting 1-2 hours. MgB$_2$ plate-like crystals up to 200~$\mu$g in
weight and $1.5\times0.9\times0.2$~mm$^3$ in size can be obtained
by using this technique, even though the crystals used in our
measurements were smaller ($0.6\times 0.6 \times 0.04 $~mm$^3$ at
most). The crystals were etched with 1\% HCl in dry ethanol to
remove possible deteriorated surface layers. The critical
temperature of the crystals, measured by AC susceptibility, is
$T\ped{c}= 38.2$ K with $\Delta T\ped{c}\sim 0.2$ K.

Using Au or Pt tips to make point contacts did not ensure
mechanical stability during thermal cycling and reproducibility of
the conductance curves. Thus, we moved to a non-conventional
technique that consists in using as a counterelectrode either a
very small ($\lesssim 50 \mu$m) drop of Ag conductive paint, or a
small piece of indium pressed on the surface of the sample. With
this technique, a control of the contact characteristics is
possible anyway, by applying short voltage pulses to the junction.
The apparent contact area is much greater than that required to
have ballistic current flow \cite{ballistic}, but the
\emph{effective} electrical contact occurs in a much smaller
region, due to the presence of parallel micro-bridges in the spot
area. On the other hand, the resistance of all our contacts was in
the range $10\div 50\, \Omega$. This, together with the estimated
mean free path for the same crystals $\ell= 80\,$nm
\cite{Sologubenko}, proves that our contacts are in the ballistic
regime. The contacts were positioned on the crystal surfaces so as
to inject the current along the $c$ axis or along the $ab$ planes.
The directionality of current injection is ensured by the small
roughness of the crystal surfaces even on a microscopic scale.
Figures 1(a) and (b) report AFM measurements on the $ab$-plane
surface, after removal of the In contact; the surfaces
perpendicular to the $ab$ plane are even smoother.

\begin{figure}[t]
\vspace{-1mm}
\includegraphics[keepaspectratio, width=\columnwidth]{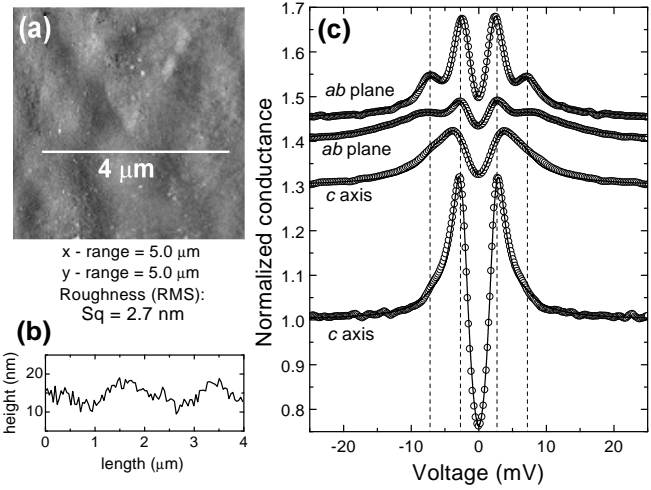}
\vspace{-7mm} \caption{(a) AFM image of the $ab$-plane crystal
surface in the contact region, after removal of the In electrode.
(b) Profile curve along the white line in (a). (c) Some
\emph{normalized} experimental conductance curves measured at low
temperature ($4.2 \div 4.6$~K) with $ab$-plane and $c$-axis
current injection. The curves are vertically displaced for clarity
and solid lines are the best-fitting curves (see text).}
\vspace{-3mm}
\end{figure}

Figure~1(c) shows some examples of the low-temperature
\emph{normalized} conductances d$I/$d$V$ of contacts with current
injection either parallel or perpendicular to the $ab$ plane. All
the conductance curves shown in the present Letter were normalized
by dividing the measured d$I/$d$V$ data by the linear or quartic
function that best fits them for $|V|\!\!>\!\!30$~meV. The
$ab$-plane curves clearly show two peaks at $V\simeq \pm 2.7$~mV
and $V\!\simeq \pm 7.2$~mV, while the $c$-axis curves only show a
peak at $V\!\simeq \pm (2.8\div 3.5)$~mV and a smooth shoulder at
$V\!\simeq \pm 7.2$~mV. These features, marked by dashed lines in
the figure, are clearly related to the two gaps $\Delta\ped{\pi}$
and $\Delta\ped{\sigma}$. Solid lines are the best-fitting curves
calculated by using the BTK model \cite{BTK} generalized to the
case of two bands, in which the normalized conductance $\sigma$ is
given by: $\sigma\!\! = \!\!w\ped{\pi}\sigma\ped{\pi}\!+\!(1\!
-\!w\ped{\pi})\sigma\ped{\sigma}$. Here, $\sigma\ped{\pi}$ and
$\sigma\ped{\sigma}$ are the normalized conductances for the $\pi$
and $\sigma$ bands, respectively, and $w\ped{\pi}$ is the weight
of the $\pi$ band, that depends on the angle $\varphi$ between the
direction of current injection and the boron planes
\cite{Brinkman}. The fit is almost perfect, especially at low
voltage, but it must be said that there are 7 adjustable
parameters: the gaps $\Delta\ped{\sigma}$ and $\Delta\ped{\pi}$,
the broadening parameters $\Gamma\ped{\sigma}$ and
$\Gamma\ped{\pi}$, the barrier height coefficients $Z\ped{\sigma}$
and $Z\ped{\pi}$, plus the weight factor $w\ped{\pi}$. The
normalization may yield additional uncertainty on
$\Gamma\ped{\sigma,\pi}$ and $Z\ped{\sigma,\pi}$ but does not
affect the gap values.

\begin{figure}[t]
\includegraphics[keepaspectratio, width=\columnwidth]{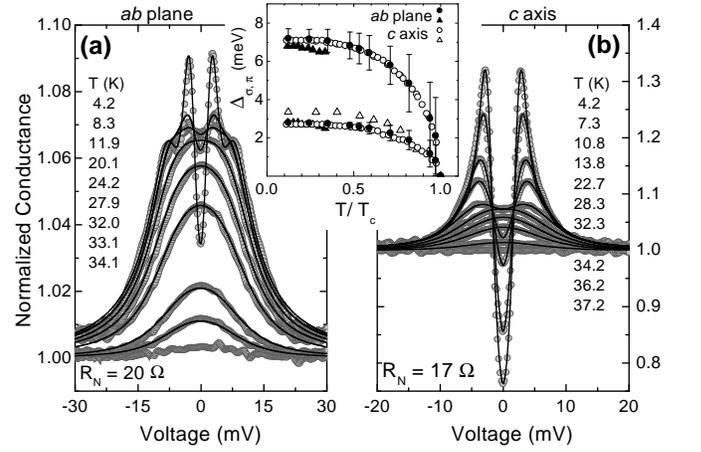}
\vspace{-6mm} \caption{Temperature dependency of the normalized
conductance in a Ag-paste contact with current parallel to the
$ab$ planes (a) and in a In-spot contact with current along the
$c$ axis (b). Solid lines are the BTK best-fitting curves. Inset:
temperature dependency of the gaps obtained from the fit of the
conductance curves of various contacts.} \vspace{-3mm}
\end{figure}

Figures 2(a) and 2(b) show the temperature dependency of the
normalized conductance curves (circles) of Ag-paint and In point
contacts, respectively. The current was mainly injected along the
$ab$ planes in (a), and parallel to the $c$ axis in (b). At the
increase of the temperature, the typical two-gap features shown in
Fig.~1 merge in a broad maximum, which disappears at the
$T\ped{c}$ of the junction that fell in all cases between 34.1 and
37.6 K. Since neither the current direction nor the contact
resistance depend on the temperature, in fitting the conductances
at various temperatures we kept both $w\ped{\pi}$ and the barrier
parameters $Z\ped{\sigma}$ and $Z\ped{\pi}$ equal to their low-$T$
values, thus reducing the actual adjustable parameters to 4. The
best-fit curves are shown in Fig.~2 as solid lines.

The inset of Fig.~2 reports the temperature dependency of the two
gaps, obtained by fitting the conductance curves of two $ab$-plane
contacts (solid symbols) and of two $c$-axis contacts (open
symbols). For clarity, error bars are only shown for a data set in
the $ab$-plane-current case, but they are of the same order of
magnitude in the $c$-axis cases. The relevant barrier parameters
(independent of temperature) are $Z\ped{\sigma}\!\!=\!\! 0.5\!\div
\!1.4$ and $Z\ped{\pi}\!\!=\!\!0.3 \!\div \!0.8$ depending on the
junction, while the broadening parameters $\Gamma\ped{\sigma,
\pi}$ increase with $T$ always remaining in the range between 0.5
and 3 meV. An important result is that the average values of the
$\pi$-band weight factor resulting from the fits
($w\ped{\pi}\!=\!0.75 \!\pm \!0.03$ for $ab$-plane current, and
$w\ped{\pi}\!=\!0.980\! \pm\! 0.005$ for $c$-axis current) are in
very good agreement with the values predicted by the two-band
model ($w\ped{\pi}\!=\!0.66$ and $w\ped{\pi}\!=\!0.99$,
respectively \cite{Brinkman}). The small mismatch that actually
exists can be ascribed to the fact that, in our low-barrier
contacts, the current is injected within a finite solid angle. The
integration of the theoretical $w\ped{\pi}(\varphi)$, taking into
account the $\cos{\varphi}$ dependency of the electron injection
probability, shows that our experimental values of $w\ped{\pi}$
are compatible with cone apertures of about $26^{\circ}$ and
$60^{\circ}$, respectively.

The average low-temperature gap values
$\Delta\ped{\sigma}\!=\!7.1\!\pm \! 0.5$~meV and
$\Delta\ped{\pi}\!=\!2.9\! \pm \! 0.3$~meV agree very well with
the theoretical values predicted by the two-band model
\cite{Choi,Brinkman}. Nevertheless, at $T/T\ped{c}\gtrsim 0.5$ the
experimental uncertainty on the gap value increases so much that
it becomes practically impossible to determine whether the
$\Delta\ped{\pi}(T)$ and $\Delta\ped{\sigma}(T)$ curves strictly
follow a BCS-like curve or not. Clearly, this problem is also
present in all the previous point-contact or tunneling experiments
in which the temperature dependency of the gaps was obtained.

\begin{figure}[t]
\includegraphics[keepaspectratio, width=\columnwidth]{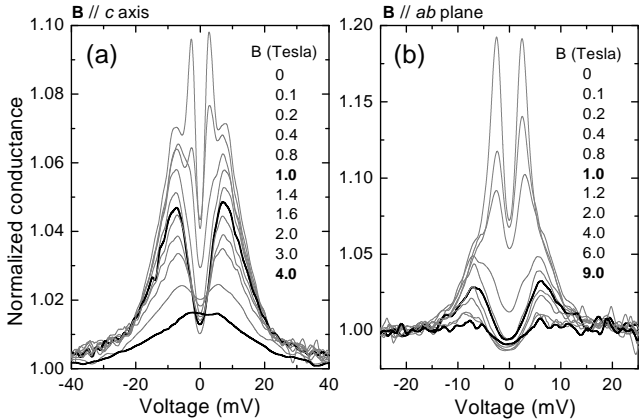}
\vspace{-7mm} \caption{(a) Some experimental normalized
conductance curves of an $ab$-plane contact, in increasing
magnetic fields parallel to the $c$ axis. Thick lines represent
the curves measured at $B=1$~T and $B=4$~T. (b) Same as in (a) but
for an $ab$-plane contact with $\mathbf{B}\parallel ab$-plane.
Thick lines represent the conductances at $B=1$~T and $B=9$~T.}
\vspace{-3mm}
\end{figure}

A careful and reliable test of the predictions of the two-band
model obviously requires a more accurate determination of the gaps
and their temperature dependency. Only by reducing the number of
free fitting parameters, e.g. by separating the contributions of
the two bands to the total conductance, this goal might be
obtained. On the basis of some point-contact results obtained  by
Szab\'{o} \emph{et al.} \cite{Szabo} in polycrystalline samples
exposed to magnetic fields, we developed a technique that combines
the selective removal of one gap with the directional
point-contact spectroscopy. By applying to each junction (at low
temperature) magnetic fields of increasing intensity, either
parallel to the $c$ axis or to the $ab$ planes, we observed in
both cases the complete vanishing of the small-gap features in the
conductance when $B\simeq 1$~T. This effect is clearly visible in
Figure~3, that shows the magnetic field-dependency at 4.2 K of the
conductance of $ab$-plane contacts in a field parallel to the $c$
axis (a) and parallel to the $ab$ plane (b). The crucial point
here is to show that a field of 1~T is enough to remove
superconductivity in the $\pi$ band without modifying the
conductance of the $\sigma$ band up to a temperature close to
$T\ped{c}$. Actually, the effect of the field on the large gap
depends on the field direction. Figure~3(b) shows that, when
$\mathbf{B}\parallel ab$ plane, the large-gap features remain
clearly distinguishable up to 9~T, with only some marks of gap
closing. Instead, when $\mathbf{B}\parallel c$ axis (a), the peaks
due to the large gap merge together at $B\geq 4$~T giving rise to
a broad maximum. In addition to this, if the field only slightly
exceeds 1~T the conductance curves remain practically unchanged
(see Fig.~3). These results demonstrate that: i) the $\pi$ band is
quite isotropic and its critical field at 4.2~K is around 1 T; ii)
the $\sigma$ band is anisotropic and, at 4.2~K,
$\Delta\ped{\sigma}$ is unaffected by a field of 1~T parallel to
the $ab$ plane; iii) the $\sigma$-band critical field parallel to
$ab$ is rather high ($> 9$ T) at low temperature, in agreement
with other results on similar samples \cite{campocritico1}. In
addition to this, some preliminary measurements we made in
$c$-axis contacts with $\mathbf{B}\parallel ab$ at about 30~K have
shown that $\mathbf{B}\ped{c2\parallel ab}^{\sigma} \sim 3.5$~T. A
detailed discussion of the temperature dependency of the critical
fields determined by our Andreev reflection experiments will be
given in a forthcoming paper \cite{nostro}. Nevertheless, on the
basis of the aforementioned results, we can be confident that a
field of 1~T parallel to the $ab$ planes is too weak to affect
seriously the large gap, even at temperatures close to $T\ped{c}$.
This hypothesis will be confirmed by the $\Delta\ped{\sigma}(T)$
curve (see Fig.~5(c)) that shows a BCS-like behavior with no
anomalous high-$T$ gap suppression due to the field.

\begin{figure}[t]
\includegraphics[keepaspectratio, width=0.9\columnwidth]{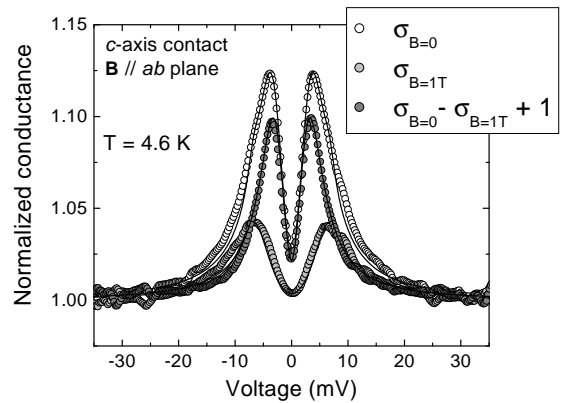}
\vspace{-2mm} \caption{Open circles: normalized conductance of a
$c$-axis contact, with no magnetic field. Light gray circles:
conductance of the same contact with a field of 1~T applied
parallel to the $ab$ plane. Dark gray circles: difference between
the two previous curves (suitably shifted). Solid lines are the
best-fit curves given by the appropriate BTK model (see text).}
\vspace{-3mm}
\end{figure}
\begin{figure}[t]
\vspace{-2mm}
\includegraphics[keepaspectratio, width=\columnwidth]{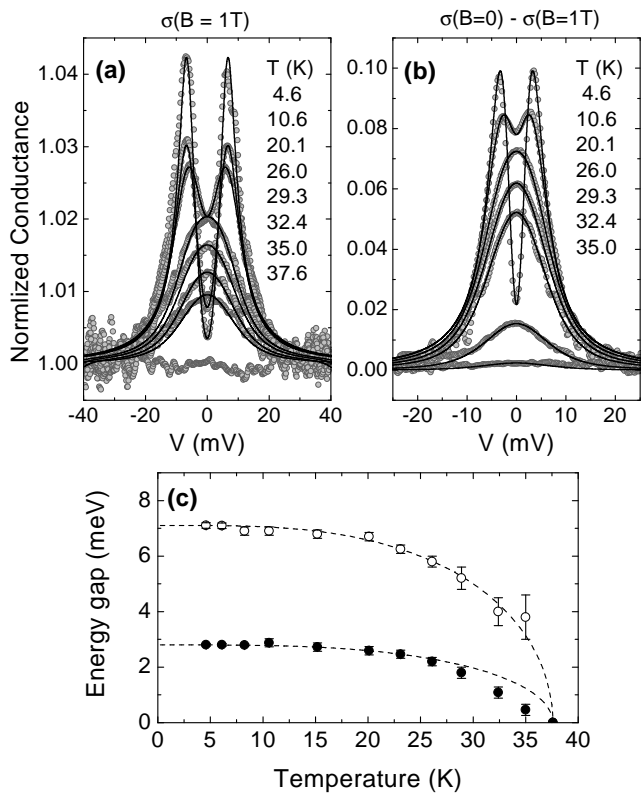}
\vspace{-20mm} \caption{(a) Temperature dependency of the
normalized conductance of a $c$-axis junction, in a field
$B_{\parallel ab}= 1$~T (symbols). (b) Temperature dependency of
the difference between the conductance in zero field (see Fig. 4)
and the conductance in a field of 1~T (previous panel). In both
(a) and (b) solid lines are the BTK best-fitting curves with three
parameters (see text). (c) $\Delta\ped{\sigma}(T)$ (open circles)
and $\Delta\ped{\pi}(T)$ (solid circles) obtained from the fit of
the curves in (a) and in (b), respectively. Dashed lines are the
corresponding BCS-like curves.} \vspace{-3mm}
\end{figure}
As a consequence, we measured the conductance of a In-MgB$\ped{2}$
$c$-axis contact at 4.6 K, with no field (see Figure 4, open
circles) and with a field of 1~T parallel to the $ab$ planes (see
Figure 4, light gray circles). When the magnetic field destroys
the gap in the $\pi$ band, the \emph{normalized} conductance
becomes: $\sigma(B\!\!=\!\!1\mathrm{T})\!=\!w\ped{\pi}\!+\!
(1\!-\!w\ped{\pi})\sigma\ped{\sigma}$. This function contains only
three free parameters: $\Delta\ped{\sigma}$, $\Gamma\ped{\sigma}$
and $Z\ped{\sigma}$, whose best-fit values at $T=4.6$~K are
7.1~meV, 1.7~meV and 0.6, respectively.  In fact, we took
$w\ped{\pi}=0.98$, that is the value obtained from the fit of the
total $c$-axis conductance at the same temperature. An independent
determination of the small gap can be obtained by subtracting the
conductance curve measured in the presence of the field from that
measured without field. The resulting curve, vertically shifted by
one unit, is reported in Fig.~4 (dark gray circles). The result of
the subtraction can be expressed by the functional form
$\sigma(B\!\!=\!\!0)\!-\!\sigma(B\!\!=\!\!1\mathrm{T})\!=\!%
w\ped{\pi}(\sigma\ped{\pi}\!-\!1)$. Fitting the experimental data
to this function (again, with $w\ped{\pi}=0.98$) allows
determining the three remaining free parameters $\Delta\ped{\pi}$,
$\Gamma\ped{\pi}$ and $Z\ped{\pi}$, that assume at $T=4.6$~K the
values 2.8 meV, 2 meV and 0.6, respectively. Incidentally, the
very good quality of the fits (solid lines in Fig. 4) further
shows that the value of $w\ped{\pi}$ is appropriate. Fig.~5
reports the temperature dependency of the curves already shown at
4.6 K in Fig.~4: the $c$-axis conductance in a field of 1~T
parallel to the $ab$ planes, $\sigma(B\!\!=\!\!1 \mathrm{T})$ (a)
and the difference $\sigma(B\!\!=\!\!0)\!-\!\sigma(B\!\!=\!\!1
\mathrm{T})$ (b), with the relevant best-fitting curves (solid
lines). Notice that the difference curves look particularly
``clean'' and noise-free since the subtraction also allows
eliminating some experimental fluctuations that are present both
in $\sigma(B\!\!=\!\!0)$ and in $\sigma(B\!\!=\!\!1\mathrm{T})$.
The resulting fits are quite good at any temperature and in the
whole voltage range. Finally, the temperature dependency of both
the large and the small gap obtained from this fitting procedure
is reported in Figure~5(c). A comparison with the inset of Fig.~2
clearly shows that the separate fitting of the partial
conductances allows a strong reduction of the error bars
(evaluated from the fitting procedure) and a consequent
improvement of the accuracy. In particular, the error affecting
$\Delta\ped{\pi}$ is very small even at $T$ close to $T\ped{c}$,
so that the deviation of the gap values from the BCS-like curve
(dashed line) results to be much larger than the experimental
uncertainty.

In conclusion, we have shown that a technique which combines
directional point-contact spectroscopy with the selective removal
of the $\pi$-band gap by a magnetic field not only proves the
existence of two gaps in MgB$\ped{2}$, but also allows a very
accurate test of the predictions of the two-band model. In
particular, by fitting the zero-field conductance curves of
directional point contacts, we obtained the weights of the
$\sigma$ and $\pi$ bands, which resulted in good agreement with
those predicted theoretically both for $c$-axis and $ab$-plane
current injection. Then, we separately analyzed the partial
conductances $\sigma\ped{\sigma}$ and $\sigma\ped{\pi}$, getting
the most accurate values of the gaps in MgB$\ped{2}$ obtained so
far: at low $T$, $\Delta\ped{\sigma}=7.1 \pm 0.1$~meV and
$\Delta\ped{\pi}=2.80 \pm 0.05$~meV. We also found that, while
$\Delta\ped{\sigma}$ follows a BCS-like temperature evolution,
$\Delta\ped{\pi}$ deviates from the BCS behavior at $T>25$~K, in
very good agreement with the two-band model. Due to the small
error affecting the gap value, this deviation is here
unquestionably determined for the first time.

This work was supported by the INFM Project PRA-UMBRA and by the
INTAS project ``Charge transport in metal-diboride thin films and
heterostructures''.\vspace{-3mm}

\end{document}